\documentclass[article,onecolumn]{aa}

\usepackage[T2A]{fontenc}
\usepackage{graphicx}

\def\bsao{{Bull. Spec. Astrophys. Obs.}}
\def\ab{{Astrophys. Bull., }}

\newcommand{\aas}{Astron. and Astrophys. Suppl. }

\newcommand{\arep}{Astronomy Reports }
\newcommand{\alet}{Astronomy Letters }

\begin{document}

\title{Instability in the system of the distant post-AGB star LS\,III+52\degr24 (IRAS\,22023+5249)}

\author{V.G.~Klochkova$^1$, A.S.~Miroshnichenko$^{2,3}$, V.E.~Panchuk$^1$,
    N.S.~Tavolzhanskaya$^1$,  M.V.~Yushkin$^1$}

\institute{1 -- Special  Astrophysical Observatory, RAS, Nizhnij Arkhyz, 369167 Russia,   \\
2 -- University of North Carolina at Greensboro, Greensboro, NC, USA, \\
3 --  Pulkovo Astronomical Observatory, Russian Academy of Sciences, St. Petersburg, Russia.
\\
   \email{Valentina.R11@yandex.ru}   }

\titlerunning{Instability of LS\,III+52\degr24 spectra}
\authorrunning{Klochkova et al }

\abstract{The optical spectra of the B-supergiant LS\,III +52\degr24 (IRAS\,22023+5249) obtained
at the 6-meter telescope with a resolution R$\ge$60000 in 2010--2021 revealed signs
of wind variability and velocity stratification in the extended atmosphere.
The  H$\alpha$ and H$\beta$ lines have a P\,Cyg type profile; their wind
absorption changes position in the range from Vr=$-270$ to $-290$\,km/s.
The intensity of the  H$\alpha$ emission reaches record values
with respect to the local continuum: $\rm I/I_{cont}\ge70$.
The stationary radial velocity according to the positions of symmetric
forbidden emissions and permitted metal emissions was taken as the
systemic velocity Vsys=$-149.6\pm0.7$km/s.  Based on the positions of
absorptions of NII and OII ions, a time variability of the radial velocity
in the range from  $-127.2$ to $-178.3$\,km/s  was found for the first time for
this star. This variability indicates the possible presence of a companion
and/or atmospheric pulsations. The change of the oxygen triplet profile
OI\,7775\,\AA{} due to the occurrence of unstable emission was registered.
The set of interstellar absorptions of the NaI\,D-lines profile in the
range from $-10.0$ to $-167.2$\,km/s is formed in the Local Arm and subsequent
arms of the Galaxy. The distance to the star, d$>$5.3 kpc, combined with
the high systemic velocity indicates that the star is located in the
 Galaxy beyond the Scutum–Crux arm. }

\maketitle

\section{Introduction}

The results of the IRAS telescope mission have opened to astronomers
the sky in infrared light. In particular, some infrared sources were
distinguished at high latitudes of the Galaxy.
Later they were identified with high luminous stars, mainly
at the evolutionary stage after the asymptotic giant branch
(further --  AGB)~\citep{Oudm1992,Oudm1996,Pottash1998,Hrivnak}.
After the optical identification of the some IRAS sources, a boom in
the study of these objects began. The results of the first decade
are presented in the well known review~\citep{Kwok}.
Some of the post-AGB supergiants are available to high resolution
spectroscopy; the reviews of spectral study performed with the
6-m telescope were published by~\citet{rev1,Envelop,rev2}.

The post-AGB stage includes far evolved stars with
initial masses in the interval 2$\div8\,M_{\sun}$. According
to~\citet{Block1995}, at the previous AGB evolutionary
stage, these stars are observed as red supergiants with an
effective temperature Teff$\approx$3000--4500\,K. The AGB
stage for stars of these masses is the final phase with
nucleosynthesis in stellar cores. The interest in AGB stars
and their closest descendants post-AGB stars is mainly
due  to the fact that the interiors of these stars in their
short-term evolution stage there are  physical conditions for
the synthesis of heavy metal nuclei and the  dredge-up of
the fresh products of nuclear reactions into the stellar
atmosphere and further into the  circumstellar and interstellar medium.
As a result of these processes, AGB stars with initial masses below
 3--4\,$M_{\sun}$ are the main suppliers (over 50\%) of all
 elements heavier than iron synthesized by the s-process,
 which consists in slow (as compared to $\beta$-decay)
 neutronization of nuclei. The details of the evolution of near-AGB
 stars and the results of  calculations of the synthesis and dredge-up
 of elements are given by~\citet{Herwig2005,Criscienzo,Liu}.

In the recent decades, a subgroup of hot supergiants has been
selected  among post-AGB stars, often with emissions in their
spectra, classified as post-AGB   stars approaching the planetary
nebula phase. A good  example is the high-latitude hot star SAO\,244567
(Teff$\ge$35000\,K), for which~\citet{Partha1993}
concluded, by comparing spectra spaced apart by 50~years, that it is
approaching the phase of a young planetary nebula.

The object of this paper is the hot supergiant LS\,III+52\degr24
associated with the infrared source IRAS\,22023+5249. In early
studies, see for example, the objects listed by~\citet{Hardorp}
mention this star among the stars emitting in H$\alpha$.  The SIMBAD
database points  the star as spectral type Be.  ~\citet{Suarez},
studying a large sample of stars with IR excesses, classified
IRAS\,22023+5249 as an object in transition to a planetary nebula.

The main features of the optical spectrum of LS\,III+52\degr24
are well known by now. ~\citet{Sarkar2012} used the
high resolution spectrum to determine the fundamental parameters
of the star and the chemical composition of its atmosphere.
Having obtained a large radial velocity from the absorption lines,
Vr=$-148.31\pm0.60$\,km/s, these authors came to the conclusion
that LS\,III+52\degr24 is an O-rich post-AGB star.
\citet{Arkh2013}  found a fast (from night to night)
variability in the UBV bands with a variability amplitude $\Delta$V=0.35\,mag.
These authors also found a correlation between the star’s brightness and
the intensity of the HI, HeI, [NII], [SII], and other lines; they
also noted an increase in the equivalent widths of the [NII] and
[SII] nebular emissions over 20~years.

    In this paper, we present the results of the optical
spectra analysis of LS III +52°24 obtained with the 6-m BTA telescope 
in 2010–2021. The main goal of our study is to search for the variability 
in the profiles of spectral features and the behavior of the radial
velocity pattern with time. The methods of observation and data analysis 
are briefly described in Section~2. In Section~3, the
results are presented in comparison with those published earlier. 
The discussion of the results and the conclusions are provided in
Section~4.

\section{Observations and data processing}

\begin{table*}[ht!]
\medskip
\caption{Results of the measurements of the heliocentric radial velocity Vr in the 
LS\,III+52\degr24 spectra from different types of lines. Columns 4--6, denoted
as (abs)/(emis), show Vr from the absorption (top) and emission (bottom)  components 
of the corresponding H$\alpha$ and HeI lines. }
\begin{tabular}{ c| c|  c|  c|  c| c }
\hline
Date/JD& \multicolumn{5}{c}{\small  Vr, km/s} \\  
\cline{2-6}
2450000+ & Absorptions&symmetric   &  H$\alpha$(abs)/  & HeI\,5876(abs)/ & HeI\,6678(abs)/  \\  
         &          &emissions      &     (emis)        &  (emis)    &  (emis) \\
\hline
   1    &   2       &  3            &   4              &  5  &  6  \\
\hline
14.07.2001$^1$&$-152.4$& $-147.3$     &  $-185.36$&   & $-182.16$  \\
          &$\pm0.3$\,(8)  & $\pm0.17$\,(15)&&   &  \\
\hline
27.09.2010 &$-178.3$ &$-149.6$       & $-272.1$ & $-228.9$ &$-210.5$ \\ 
 5467.43   &$\pm0.2$\,(9) &$\pm$0.08\,(29)& $-129.7$ & $-125.1$ &$-121.2$  \\    
\hline
07.12.2019 &$-151.3$ &$-150.3$       & $-274.6$ &$-229.2$  &$-210.0$  \\  
 8825.23   &$\pm0.3$\,(11)&$\pm$0.06\,(37)& $-138.2$ &$-131.3$  &$-125.6$ \\   
\hline
 29.08.2020&$-140.8$& $-150.0$ & $-290.5:$&$-210.7$  &$-201.1$   \\    
9091.47    &$\pm0.2$\,(12)&$\pm$0.07\,(33)  & $-149.6$ &$-121.7$  &$-122.5$  \\
\hline
26.10.2020 &$-127.2$& $-148.6$       & $-271.3$ &$-213.3$  &$-201.0$ \\ 
9149.27   &$\pm0.14$\,(15) & $\pm$0.06\,(46)& $-149.5$ &$-122.6$&$-123.2$ \\
\hline
29.07.2021 & $-141.9$&$-150.1$  &$-273.1$ &$-229.4$ & $-217.6$ \\  
9424.52    & $\pm0.4$\,(7) &$\pm$0.06\,(32) &$-148.9$&$-123.8$ &$-126.5$\\
\hline
\multicolumn{6}{l}{\footnotesize 1 --  Vr values for 2001 year are obtained by averaging  the corresponding data
by~\citet{Sarkar2012}}\\
\end{tabular}
\label{velocity}
\end{table*}

    The spectra of LS\,III+52\degr24 were obtained with the NES echelle
spectrograph~\citep{NES} positioned at the Nasmyth focus of the 6-meter  telescope BTA.
The dates of the observations of the star are listed in Table~\ref{velocity}. 
The NES echelle spectrograph is equipped with a large-format CCD  
with 4608\,×\,2048 elements and an element size of 0.0135\,×\,0.0135 mm; 
the readout noise is 1.8e$^-$. The registered spectral range is
$\Delta\lambda$\,=470--778\,nm. To reduce light losses without loss
of spectral resolution, the NES spectrograph is equipped with an image 
slicer that splits the image into three slices. Each spectral order 
in the 2-dimensional image of the spectrum is repeated three
times. The spectral resolution is $\lambda/\Delta\lambda\ge$60000,
the S/N ratio along the echelle order in the spectra of LS\,III+52\degr24
varies from 40 to 60.
     One-dimensional data were extracted from two-dimensional echelle spectra using a
     modified (considering the features of the echelle frames of the spectrograph) ECHELLE
     context of the MIDAS software package. The details of the procedure are described
     by~\citet{MIDAS}.  The traces of cosmic
 particles were removed by median averaging of two
 spectra obtained sequentially one after the other. The
 wavelength calibration was carried out using the spectra of a Th–Ar 
lamp with a hollow cathode. All further etapes of 
 processing were performed using the last version of the DECH20t 
code developed by G.~Galazutdinov. The systematic errors in measuring 
heliocentric velocities Vr estimated from sharp interstellar NaI
 components and telluric lines do not exceed 0.25\,km/s
 (for one line); the random errors for shallow absorptions are
 $\approx$0.5\,km/s (the average value per one line).
 For the average Vr values in Table\,\ref{velocity}, the errors are
 0.06--0.3\,km/s depending on the number of lines measured. 
The features in the  LS\,III+52\degr24 spectrum
 were identified using the data of~\citet{IRAS01005a}
 from the spectral atlas for a hot post-AGB star associated with the IR 
source IRAS\,01005+7910. In addition, we used the results of identifying 
features in spectra of related objects from the studies of
 \citet{Sarkar2012, Sarkar2005}. To refine individual data for spectral lines,
 we also used the data from the VALD database~\citep{VALD1, VALD2}.

\section{Results}

\subsection{Variability of the optical spectrum of LS\,III+52\degr24  and radial velocity patterns}

\begin{figure*}[ht!]
\includegraphics[angle=0,width=0.55\textwidth,bb=25 80 550 680,clip]{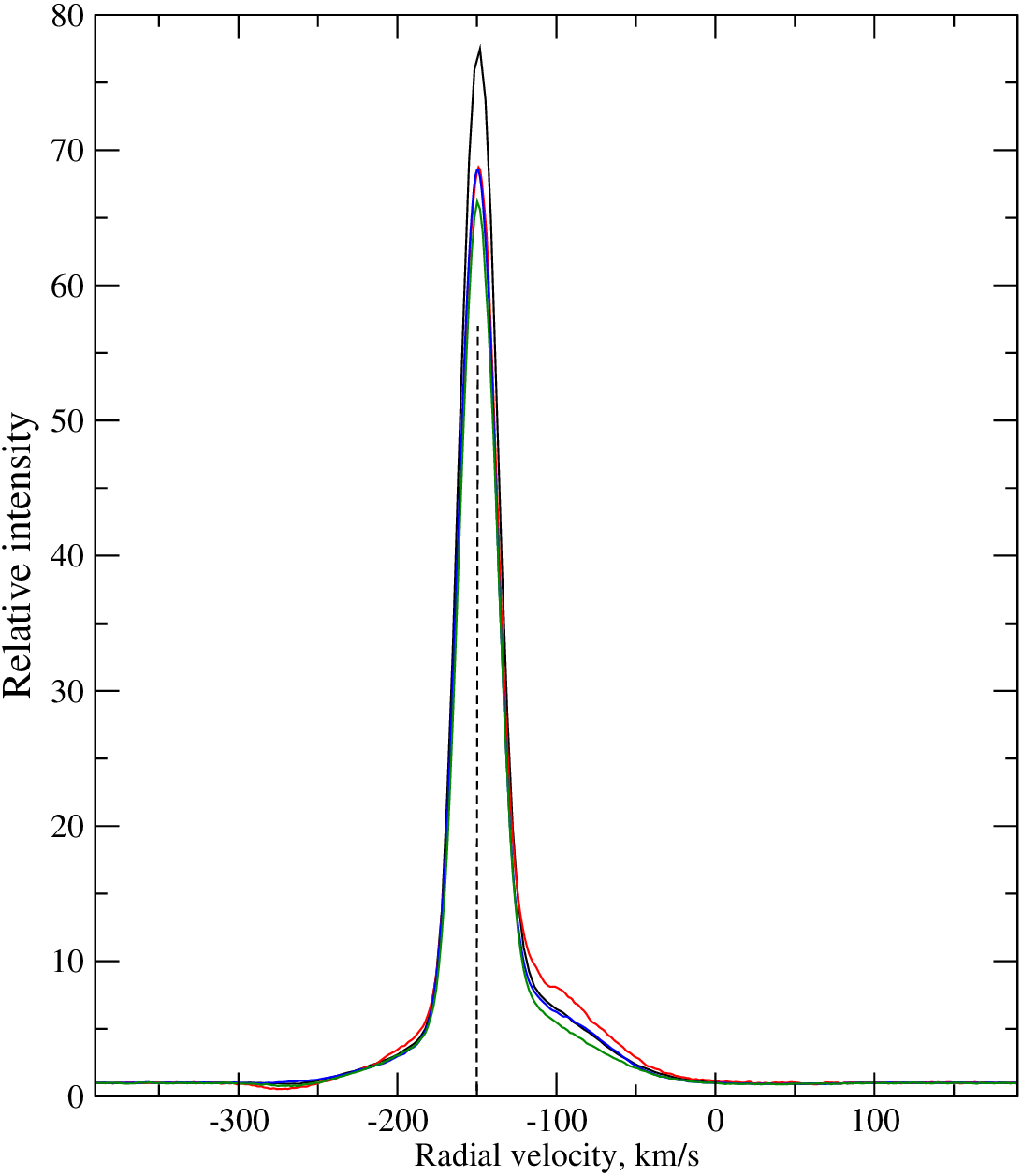}
\caption{H$\alpha$ profile in the coordinates of radial velocity versus relative intensity
in the  LS\,III+52\degr24 spectra obtained on September~27, 2010 (red line),
December~7, 2019 (green line), August~29, 2020 (blue line), and October~26,
2020 (black line). Here and below,  the position of the dashed vertical line
coincides with the accepted value of the system velocity Vsys=$-149.6$\,km/s.}
\label{Halpha}
\end{figure*}

\begin{figure*}[ht!]
\includegraphics[angle=0,width=0.6\textwidth,bb=10 80 555 680,clip]{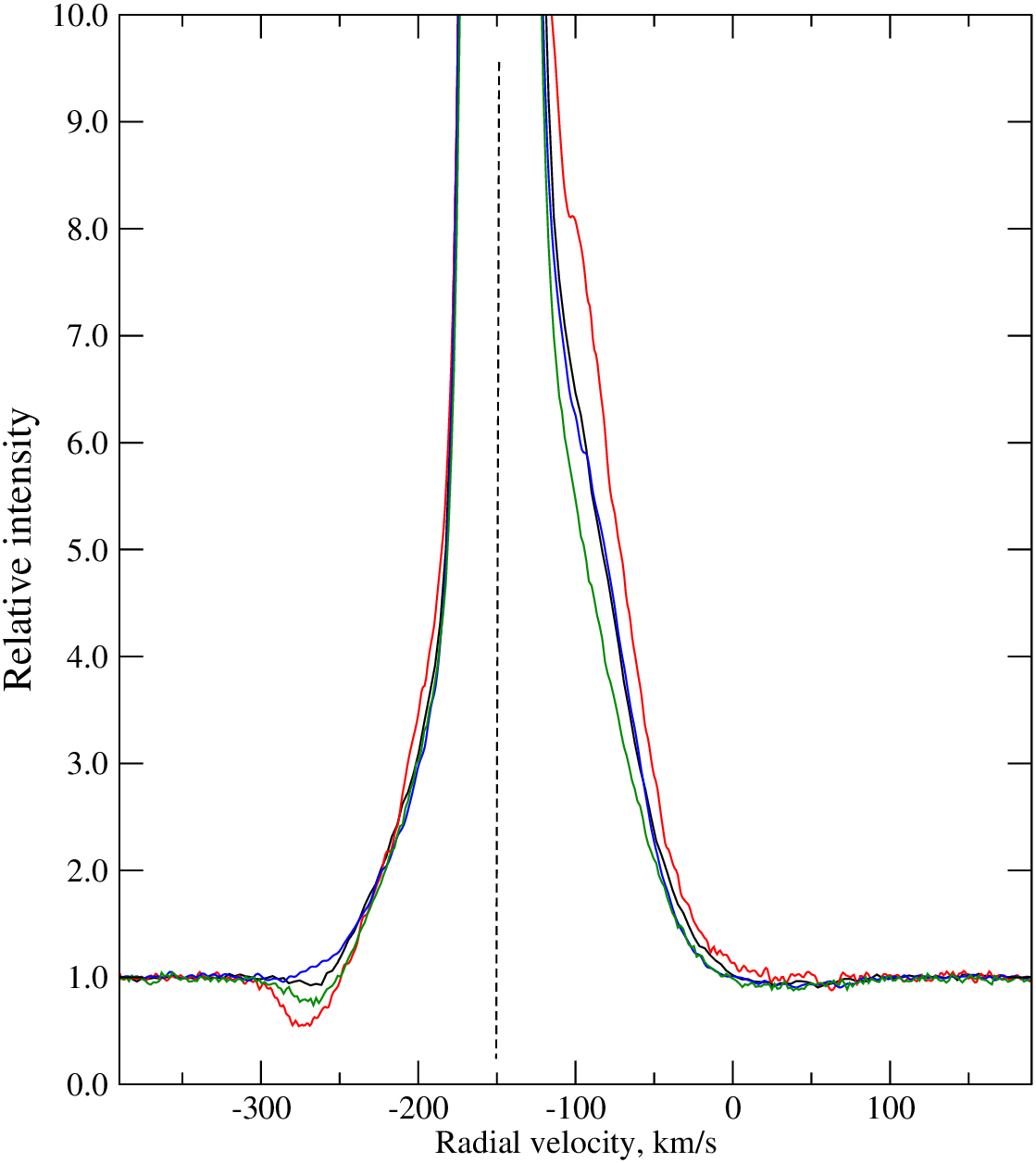}
\caption{Same as in Fig.~1 for the lower part of the H$\alpha$ profile}
\label{Halpha_low}
\end{figure*}

\begin{figure*}[ht!]
\includegraphics[angle=0,width=0.55\textwidth,bb=15 75 550 675,clip]{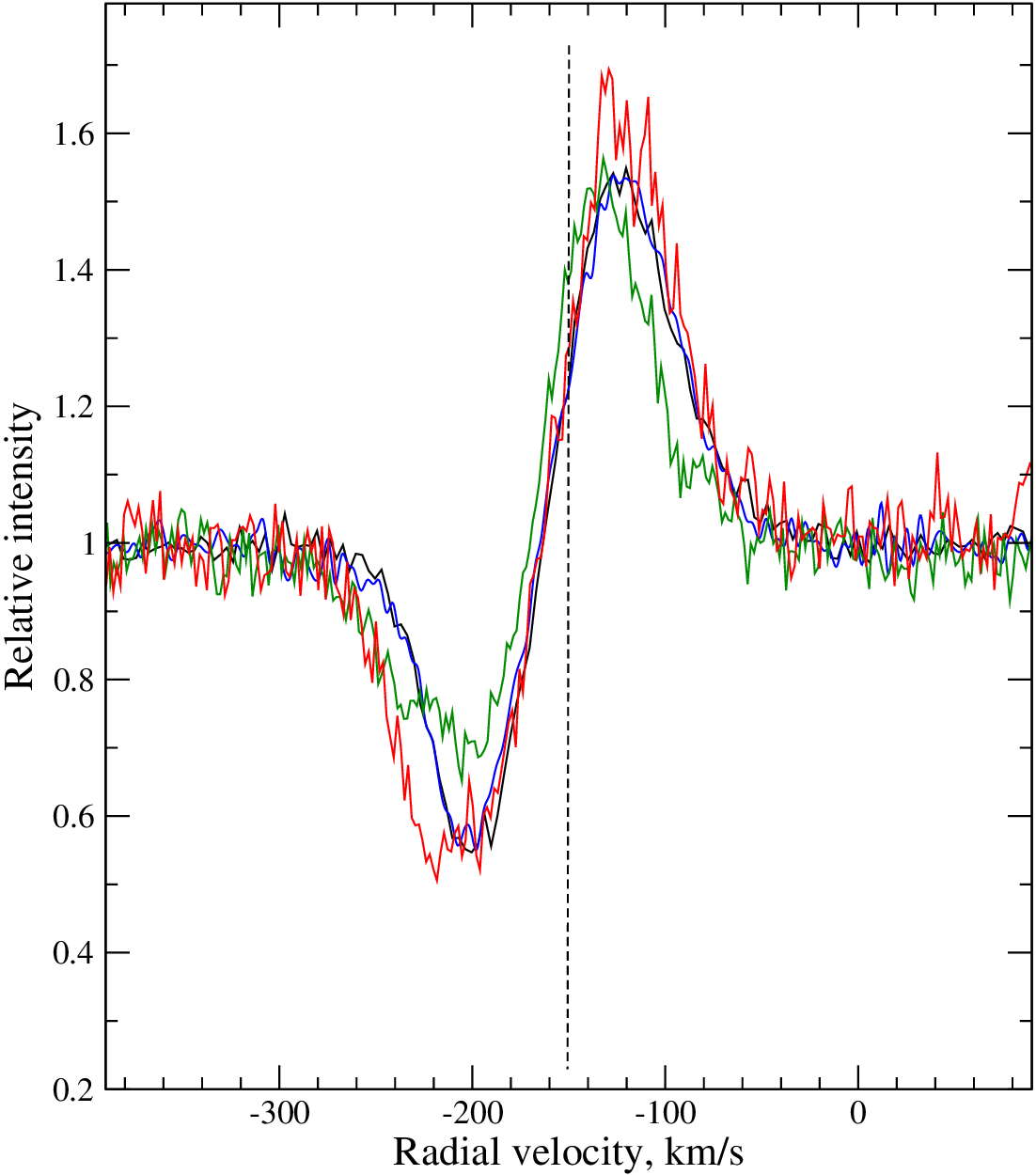}
\caption{Same as in Fig.\,1 for the HeI\,6678\,\AA{} line.}
\label{He6678}
\end{figure*}

\begin{figure*}[ht!]
\includegraphics[angle=0,width=0.55\textwidth,bb=15 75 550 675,clip]{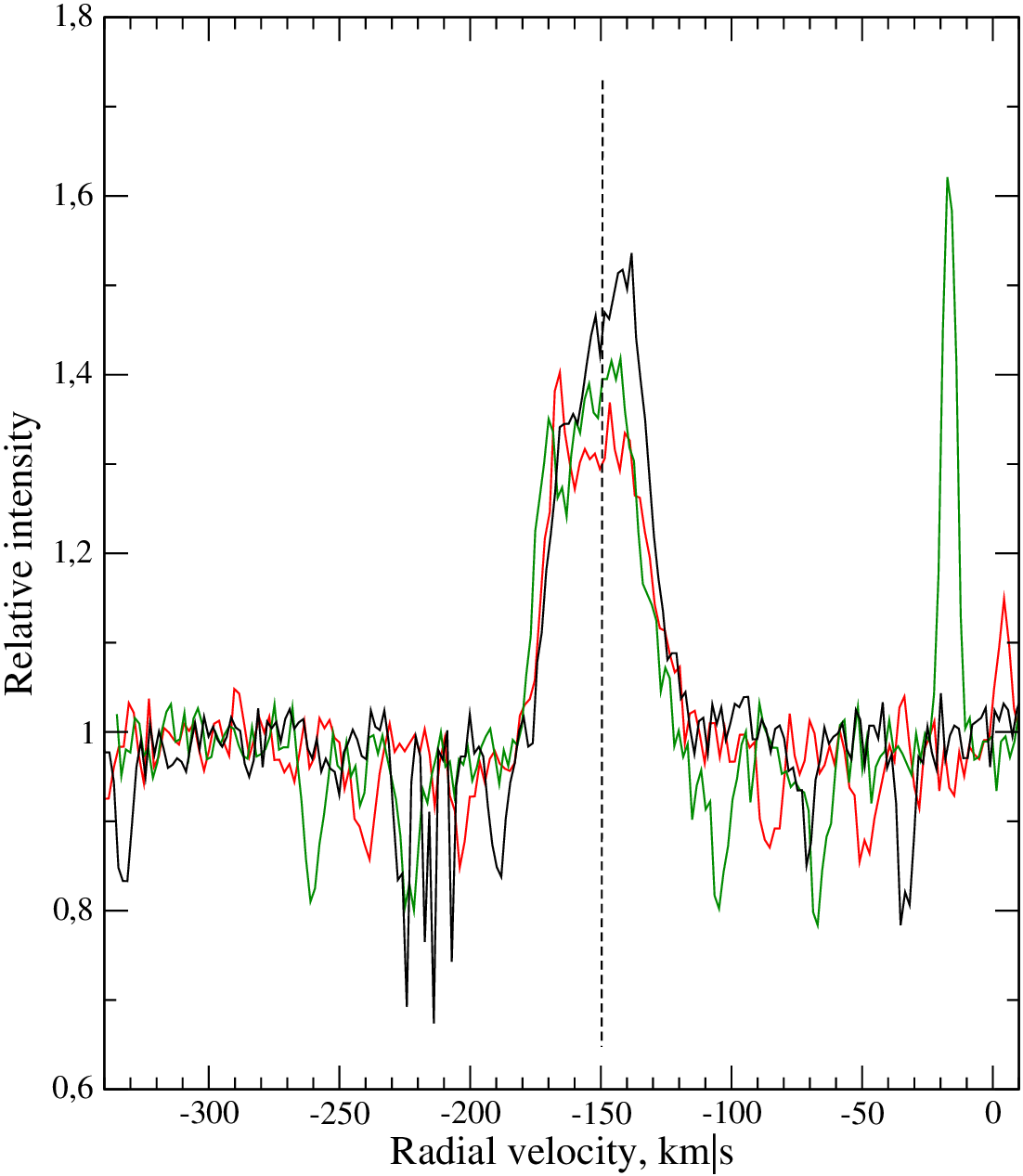}
\caption{Profile of the [OI] 6300 Å line in the LS\,III+52\degr24 spectra taken on
September~27, 2010 (red line), December~7, 2019   (green line), and July~29, 2021 (black line).
The narrow emissions in this fragment are the [OI]\,6300\,\AA{} telluric line }
\label{6300}
\end{figure*}

     The main features of the optical spectrum of hot
 post-AGB stars are currently known fairly well. Let us
 refer to the results of high-resolution spectroscopy
 published by~\citet{Garcia1997}, \citet{IRAS01005a}, \citet{Sarkar2012, Sarkar2005},
\citet{Mello2012}, and~\citet{Ikon2020}. The optical spectrum of LS\,III+52\degr24
 is a composition of the spectrum of a hot
 supergiant and the emission-rich spectrum of a circumstellar nebula. 
The spectra contain three types of  emissions: the emission component of the complex
 lines of neutral hydrogen and helium, as well as
 numerous symmetrical permitted (OI, SiIII, AlIII, CII,
 FeI, FeII, FeIII) and forbidden low-excitation emissions ([NII], [OI], [SII]). 
The occurrence of forbidden emissions [NII], [SII] indicates the approach to
 the planetary nebula phase. The profiles of each of
 these types of lines are presented in Figs.~\ref{Halpha}, \ref{Halpha_low}, \ref{He6678},
 \ref{6300}.
     All forbidden emissions in the [NII] spectrum have the simplest symmetric profile: 
a narrow Gaussian with a half-width of $\approx$10\,km/s. The profiles of
forbidden oxygen emissions are more complex.

 As follows from Fig.\,\ref{6300}, the slopes of the [OI]\,6300\,\AA{} emission
profile are almost vertical and are approximately $\pm20$\,km/s away 
from the center of the profile. The half-width of the profile of the 
same emission, but of a telluric origin, is many times lower than $\approx$3\,km/s. 
The [OI]\,6300\,\AA{} emission profiles presented in Fig.\,\ref{6300} for
three observation  dates allow us to note the variability of this line, 
which may reflect the complex structure of the gas envelope of the star.

    Table\,\ref{velocity} contains the measurement results of the 
heliocentric radial velocity Vr in the LS\,III+52\degr24 spectra
from the positions of the samples of various types of lines: absorptions, 
symmetric forbidden and permitted emissions, emission and absorption components of
H$\alpha$ and HeI lines. The numbers in parentheses indicate the number of 
features used in the averaging. As follows from the data in the table, 
the velocity from the sample of symmetric emissions formed in the 
circumstellar envelope does not change for all observation dates.
The constancy of this value allows us to take its average as the system 
velocity of LS\,III+52\degr24:  Vsys=$-149.6$\,km/s.

    Multiple observations over a long time period allow us to discover 
the variability of the positions of pure ion absorptions. As follows from 
the data in the 2-d column of the table, the average velocity over the 
sample of absorptions of OII and NII ions varies in the
range from $-127.2$ to $-178.3$\,km/s, which is a manifestation of 
instability in the deep layers of the  stellar atmosphere. This variability 
may be due to the  pulsations in the extended atmosphere of the 
supergiant or the presence of a companion in the system.

\begin{figure*}[ht!]
\includegraphics[angle=0,width=0.5\textwidth,bb=15 75 550 675,clip]{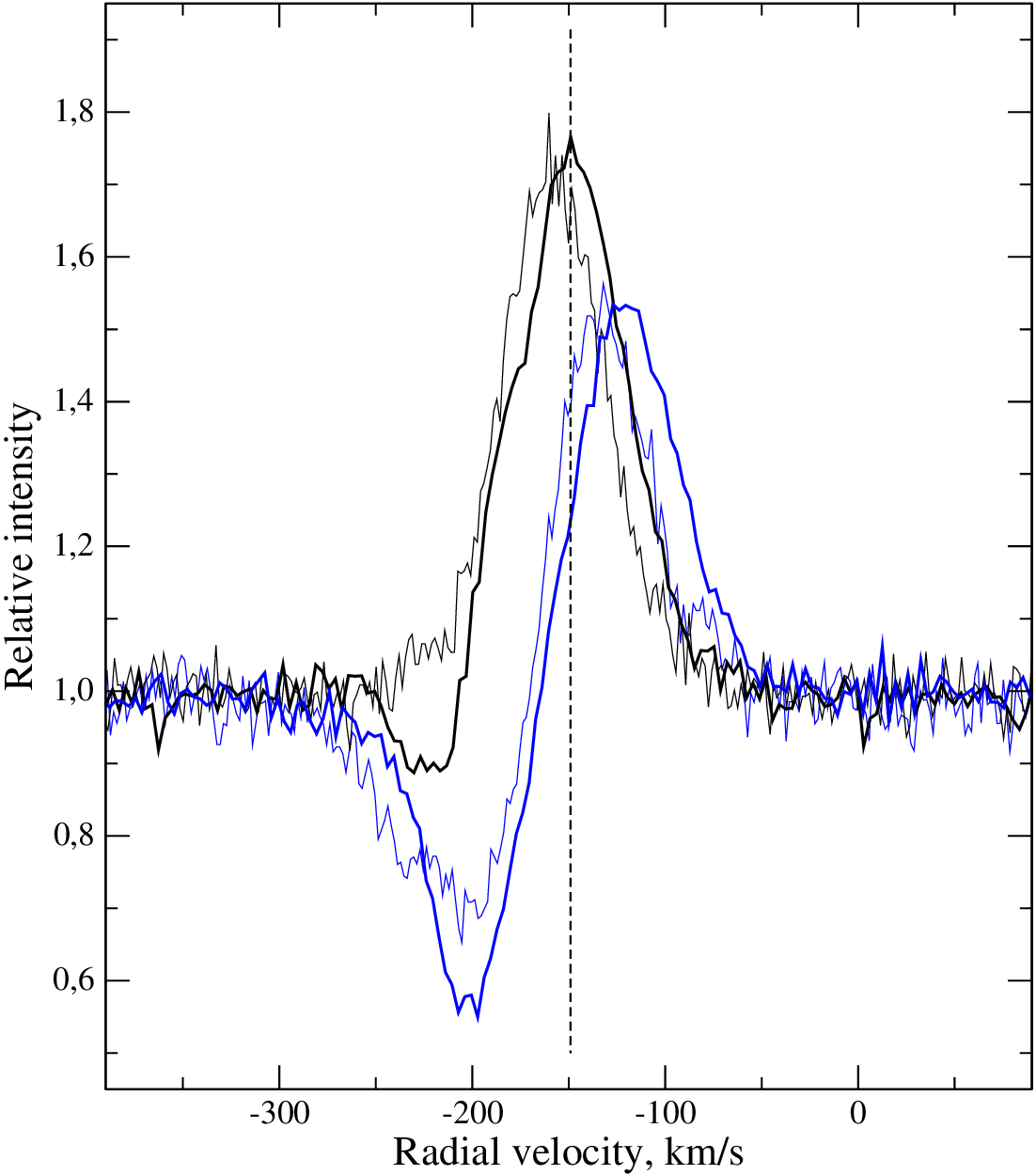}
\caption{Variability of the HeI\,6678 (blue lines) and HeI\,7065\,\AA{} (black lines)
line profiles in the LS\,III+52\degr24 spectra taken on December~7, 2019 (thin lines)
and August~29, 2020 (thick lines). }
\label{He_2lines}
\end{figure*}

       The variability in the intensity of the H$\alpha$ and HeI
  lines was discovered earlier by~\citet{Arkh2013}
  from low-resolution spectra. Our observations allow
  us to refine this result. The H$\alpha$ line profile shown in
  Fig.\,\ref{Halpha} in the coordinates of relative intensity versus
  radial velocity contains a strong emission for all the
  observation dates. We emphasize that LS\,III+52\degr24
  holds the record for the emission power in H$\alpha$: as follows
  from Fig.\,\ref{Halpha}, the H$\alpha$ emission
intensity   with respect to the local continuum reaches
  $\rm I/I_{cont}\ge 70\div78$. The position of this emission does
  not change with time and coincides with the system
  velocity Vsys=$-149.6$\,km/s we adopted. The variability of the 
emission intensity in H$\alpha$ indicates the variability of the stellar 
wind power and the inhomogeneity of the star’s gas envelope.

       Figure\,\ref{Halpha_low}, which shows the lower fragment of the
  H$\alpha$ profile, illustrates the shift in the position of the
  absorption component and the change in the depth of this wind feature, 
 which forms in the upper layers of the outflowing atmosphere at the 
base of the stellar  wind. From the data in the 4-th column of Table\,\ref{velocity},
  the variability range of the wind absorption position is
  from $-270$ to $-290$\,km/s. The terminal velocity
  reaches $-300$\,km/s. The fragment of the H$\alpha$ profile in 
Fig.\,\ref{Halpha_low} clearly shows the presence of a variable
additional emission component in the long-wave wing of the H$\alpha$ profile.

    The LS\,III+52\degr24 spectrum also indicates a
significant date-to-date variability of neutral helium lines
with profiles of the P\,Cyg type. To illustrate this phenomenon, 
Fig.\,\ref{He6678} compares the profiles of the HeI\,6678\,\AA{} 
line in the spectra for three observation dates. Here, we can 
clearly see the variability of the intensity and position of 
the emission and absorption components. In this case, the terminal velocity
reaches the same values as on the H$\alpha$ profile. Additionally, 
Fig.\,\ref{He_2lines} shows the profiles of two HeI\,6678 and
7065\,\AA{} lines for two dates of our observations: December~17, 2019 and 
August~29, 2020. The most interesting and new detail here is the enhancement 
of wind absorption in HeI\,6678\,\AA; at the same time, in August~2020, 
this kind of wind absorption formed for the first time near the 
HeI\,7065\,\AA{} line.

\begin{figure*}[ht!]
\includegraphics[angle=0,width=0.5\textwidth,bb=15 75 550 675,clip]{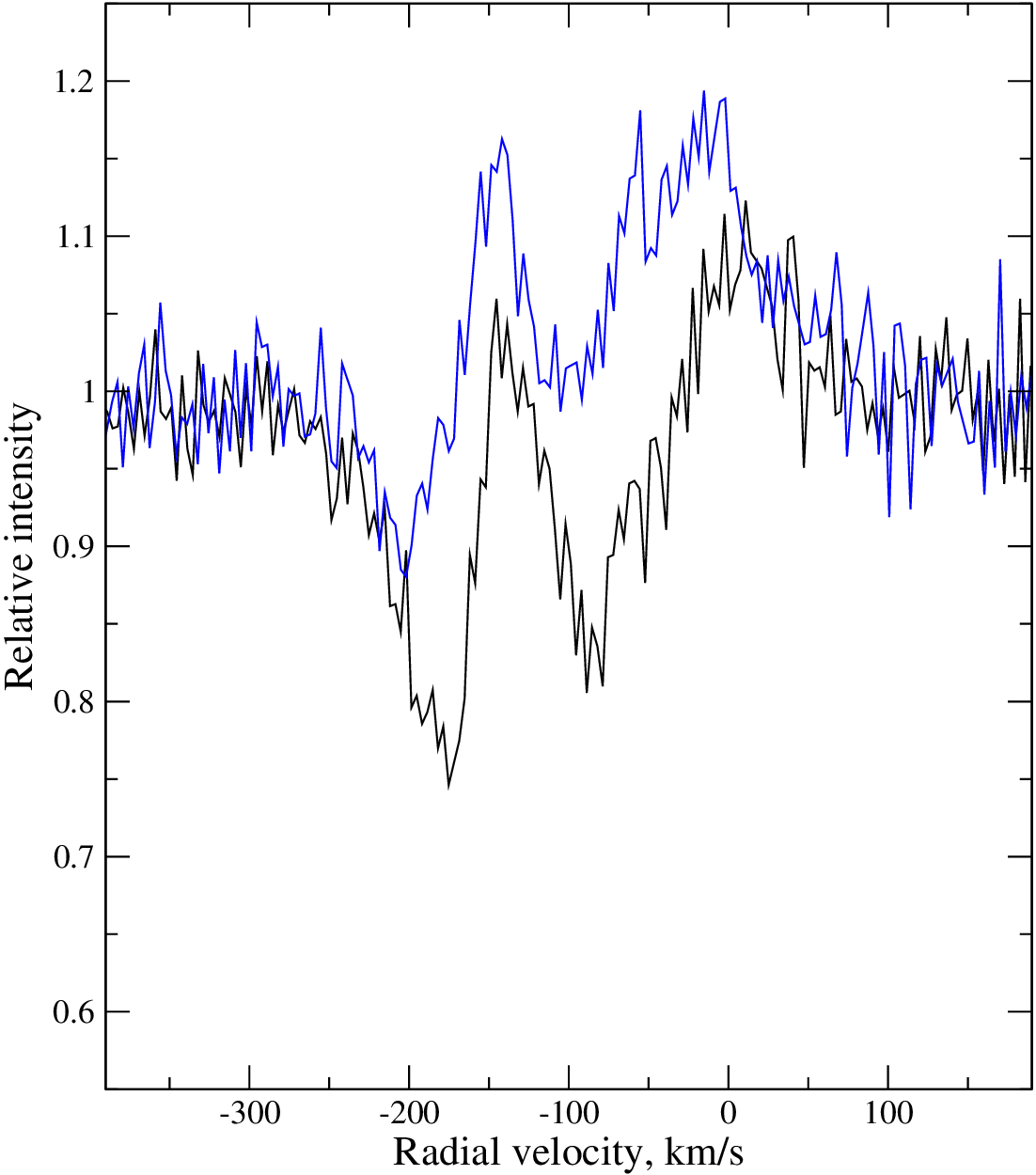}
\caption{Profile of the OI\,7775\,\AA{} triplet in the LS\,III+52\degr24 spectra
taken on August~29, 2020 (blue line) and October~26, 2020 (black line).}
\label{O_triplet}
\end{figure*}

In addition, a rare feature was recorded in the spectrum of LS\,III+52\degr24:
a significant variability in the profile of the infrared oxygen triplet, 
OI\,7775\,\AA{}, which is illustrated in Fig.\,\ref{O_triplet}, where 
the triplet profiles for two observation dates are compared. This feature,
along with the emission components in the NaI D-lines profile, was 
mentioned in passing in the paper by~\citet{Mello2012} dedicated
to the spectroscopy of hot post-AGB stars. In addition, \citet{Arkh2013},
give in their Table\,8 two values of the total equivalent width
of OI\,7775\,\AA{} emission for the oxygen triplet, which
also indicates the variability of the triplet profile.

     We also note that the forbidden emissions [SII]\,6717 and 6731\,\AA{} 
are systematically shifted to the short-wave region by approximately $-20$\,km/s
relative to other forbidden emissions. This feature, due to the
stratification of the gas envelope, is preserved in our spectra from date to date. 
It was previously noted by~\citet{Sarkar2012} for its 2001 spectrum  as well.

\subsection{Distance to the star and its Luminosity}

\begin{figure*}[ht!]
\includegraphics[angle=0,width=0.5\textwidth,bb=15 75 550 675,clip]{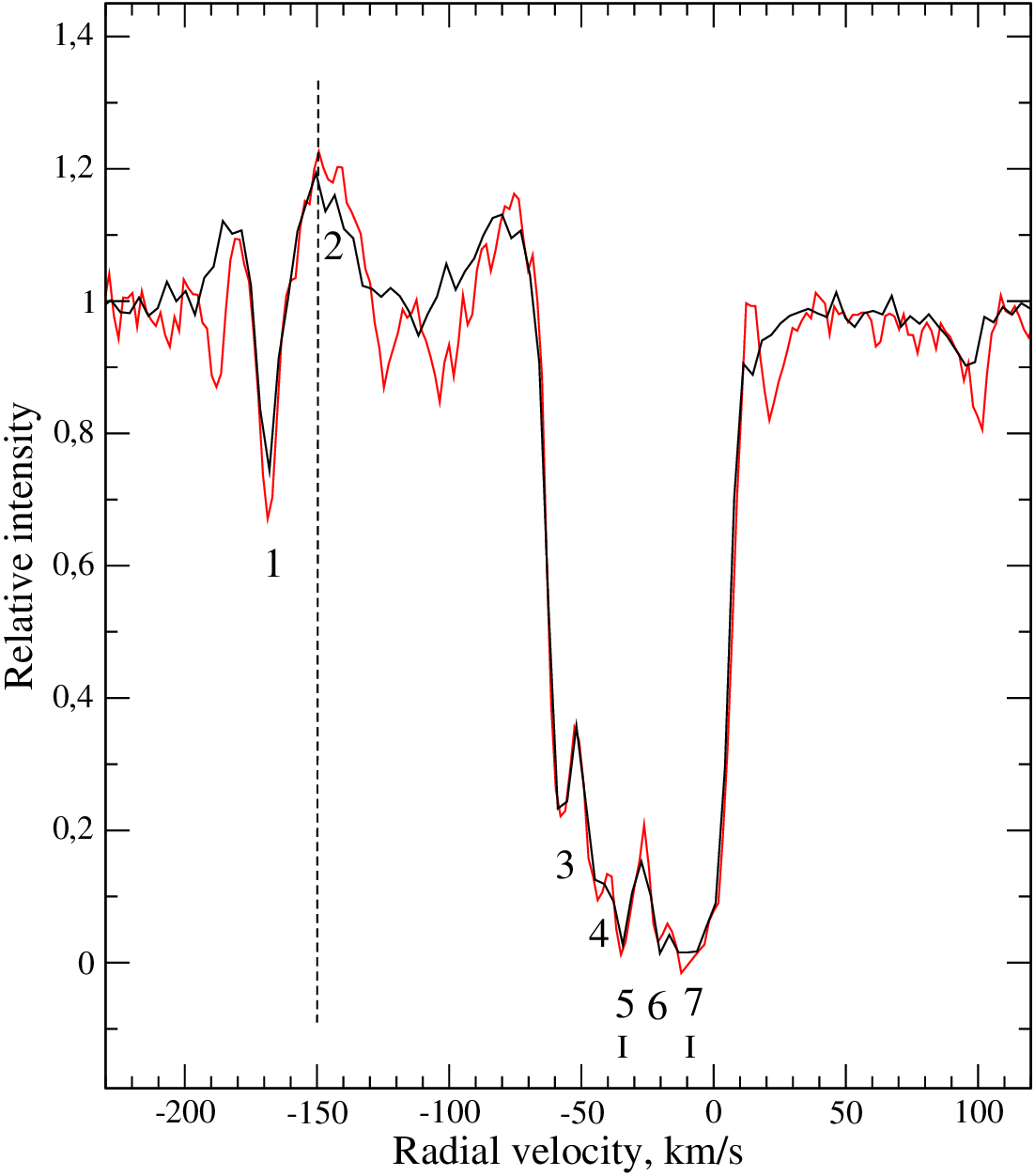}
\caption{Multicomponent profile of the NaI\,5889\,\AA{} line in the LS\,III+52\degr24
spectra taken on August~29, 2020 (black line) and October~26, 2020 (red line).
Two short verticals correspond to the velocities of the interstellar components of the
KI\,7696\,\AA{} line.}
\label{Na5889}
\end{figure*}

    The parallax of LS\,III+52\degr24 from the Gaia~EDR3 catalog,
which is measured with high accuracy  ($\pi=0.17313\pm 0.018$\,mas), 
leads to a large distance to the star: d=5.84$\pm$0.6\,kpc. 
The SIMBAD database shows the parallax from Gaia~DR2, which has too
low accuracy, $\pi=0.0804\pm0.0524$\,mas. \citet{Bailer} refined the parallax
$\pi$=0.19\,mas  on the basis of modeled Gaia~DR3 data, and  given the corresponding
distance d\,=\,5.34\,kpc.
    The significant distance to the star is confirmed by the presence of 
interstellar components that do not belong to the Local Arm in the 
structure of the   D-lines profile of the NaI doublet. 
The multicomponent profile of the NaI\,5889\,\AA{} doublet 
for two observation dates is shown in Fig.\,\ref{Na5889}.
Here, the numbers   indicate the components that form in different 
layers  of the circumstellar and interstellar medium. The
  short vertical lines in this figure indicate the positions
  of the two interstellar components of the KI\,7696\,\AA{}   line. 
Absorption components 3--7 in the range of velocities from  $-10.4$ to 
$-56.1$\,km/s are of interstellar origin. Emission~2 forms in the 
circumstellar  gaseous medium, and its position $\approx-150.2$\,km/s 
is  consistent with the systemic velocity.

The absorption components  of the NaI doublet line profiles  (from $-10$ to $-72$\,km/s) 
were  previously recorded by~\citet{IRAS01005b} in the spectrum
of the central  star of the related object IRAS\,01005+7910 located above the galactic 
plane (its latitude b=+16.6\degr).
      The shortwave absorption ``1'', the position of which,
 Vr=$-170$\,km/s, does not change from date to date of observations, 
probably forms in the circumstellar envelope expanding at a rate of  approximately
Vexp=$-20$\,km/s. This estimate of the envelope expansion velocity does not
contradict the values of this parameter from the paper by~\citet{Sarkar2012},
who estimated the expansion rate based on the widths of forbidden lines [NII] and [SII]. 
The expansion rate from the [OI]\,6300 and 6363,\AA{} forbidden line profiles 
substantially exceeds this parameter, which we also see from
our observations. The profile of the [OI]\,6300\,\AA{} line shown in Fig.\,\ref{6300} 
is repeatedly wider and, possibly, structured.

It should be noted that the interstellar component Vr$\approx -12$\,km/s 
forming in the Local Arm of the Galaxy was discovered earlier by 
\citet{IRAS22223}  in the spectrum of the post-AGB star V448\,Lac (=IRAS\,22223+4327).
This star has galactic coordinates close to those of LS\,III+52\degr24,
but a larger parallax, $\pi=0.2375\pm0.0670$\,mas; this corresponds to a 
distance of approximately 4.2\,kpc, which is consistent with the distance 
based on the V448\,Lac system velocity from radio observations.

     The average velocity for the diffuse interstellar  bands (DIBs) identified 
in the available LS\,III+52\degr24   spectra, Vr(DIBs)=$-16.0\pm0.2$\,km/s,
is consistent  with the velocity for the NaI and KI interstellar components. 
To estimate the interstellar extinction, we  used the equivalent widths W$_{\lambda}$ 
of the DIBs available in our spectra and the ratios between the color excess,
 E(B-V), and  W$_{\lambda}$ according to the calibrations from~\citep{Kos}.
Table\,\ref{DIBs} shows the   W$_{\lambda}$  values averaged  over our
spectra, and the corresponding color excesses  are given in the last column. 
For two lines absent in the  publication by~\citet{Kos}, the E(B-V) values in
italics were obtained using the calibration   dependences from~\citet{Luna}.
The average over eight DIBs E(B-V)=0.33\,mag. This estimate of reddening
 agrees well with the interstellar reddening from~\citep{Green}
near the Galactic plane in the direction of the Scutum–Crux arm.

     The estimate of reddening E(B-V) we obtained for LS\,III+52\degr24
is two times lower than the value E(B-V)=0.66\,mag from~\citep{Arkh2013}.
Such a significant difference is due to the difference in methods for 
assessing the reddening. In our case, the estimate was made on the basis 
of the equivalent widths of interstellar bands measured in the spectrum, and
in~\citet{Arkh2013}  the reddening was determined by comparing the
observed colors (U-B) and (B-V) with the normal colors of standard supergiants 
of the corresponding spectral class. Thus, an estimate was obtained for the
total color excess due to the total absorption in the interstellar medium and 
the circumstellar envelope of the star. Such a significant difference in 
color excess due to interstellar and total extinction is a typical property 
for post-AGB stars (see the sample of related post-AGB stars in the paper
by~\citet{Gauba}  as an example).

\begin{table*}[ht!]
\medskip
\caption{Equivalent widths of DIBs in the  LS\,III\,+52$\degr$24 spectra.
 The third column indicates the corresponding color excesses
obtained using the calibrations by~\citet{Kos}; the values according to the
calibrations by~\citet{Luna} are italicized.}
\begin{tabular}{c| r c }
\hline
 $\lambda$,& W$_{\lambda}$, & E(B-V) \\
 \AA{}   &  m\AA{} & mag \\
\hline
5780.48 &  334  &  0.56   \\
5797.06 &   58  &  0.29   \\
6195.98 &   22  &  0.33   \\
6283.84 &  496  &{\it 0.55}\\
6379.32 &   17  &  0.15   \\
6613.62 &   62  &  0.24   \\
6660.71 &   19  &  0.40   \\
7224.03 &  132  &{\it 0.53}\\
\hline
\end{tabular}
\label{DIBs}
\end{table*}

   Using the standard ratio of the total absorption to
the color excess, R\,=\,3.2, and the color excess from the
data of~\citet{Arkh2013}, we obtain the total absorption for LS\,III+52\degr24
Av=2.11\,mag. Having a reliable distance to the star (d=5.8\,kpc) and total extinction,
as well as using the effective temperature Teff=24000\,K published by~\citep{Sarkar2012}
and the bolometric correction BC$_V=-2.5$, corresponding to this temperature,
we can estimate the bolometric magnitude Mbol=$-5.9$\,mag and luminosity of 
the star log\,L/L$_\odot$=4.27. The parallax-based luminosity from~\citet{Bailer}
is slightly lower: log\,L/L$_\odot$=4.18. Taking into account the 1000\,K
uncertainty of the effective temperature~\citep{Sarkar2012}, we  arrive at the average
value of the luminosity   log\,L/L$_\odot=4.2\pm 0.3$.

 The data from~\citep{Partha2020} indicate that the
luminosity of LS\,III+52\degr24  obtained by us is typical of a post-AGB star.
It should be taken into account that~\citet{Partha2020}
estimated the  parameters for the post-AGB sample using the star parallaxes from 
Gaia~DR2 and gave only the lower bound for the luminosity estimate for 
LS\,III+52\degr24.
    It should be  noted that the real luminosity of LS\,III+52\degr24
 may be slightly lower if we consider that the apparent brightness of the 
star is enhanced due to the presence of powerful emissions in its spectrum. 

The luminosity we obtained log\,L/L$_\odot=4.2\pm 0.3$ serves as an additional 
indication that the star does not belong  to the supergiants with the B[e] 
phenomenon, the luminosity of  which is much higher. According to~\citet{Mir2007},
the average luminosity for a sample of this type of supergiants in the Galaxy is 
higher: log\,L/L$_\odot=5.1\pm0.2$.

\section{Discussion  of the results and conclusions}

The observed variability of the absorption--emission profiles of HI and HeI 
lines in the spectrum of LS\,III+52\degr24 indicates the inhomogeneity and the
presence of structure in its circumstellar gaseous envelope. This inhomogeneity 
and the absence of spherical symmetry are also recorded in the near-IR
images of IRAS\,22023+5249 obtained by~\citet{Gledhill}
with the NIFS instrument of the 8.2-meter Gemini North telescope. The spectrum 
of the extended gaseous envelope contains a rich spectrum of molecular hydrogen; 
the HI and HeI emissions have profiles of the P\,Cyg type. The image in the Br\,$\gamma$ 
band is especially informative, where, as \citet{Gledhill}  emphasize,
there is a strong central emission peak, an additional elliptical envelope, as 
well as bright spots and curved features. Such a structured circumstellar
medium can also explain the occurrence and variability of additional emission 
in the longwave wing of the H$\alpha$ profile in our Fig.\,\ref{Halpha_low}.

An unexpected property of the supergiant LS\,III+52\degr24 is its high system
velocity, Vsys=$-149.6\pm0.7$\,km/s. This  feature is the decisive argument for the
fact that the star belongs to the old population of the  Galaxy. This high 
velocity is consistent with the large remoteness of the star, d$>$5.3\,kpc, obtained
from its rather reliable parallax according to the Gaia~DR3 data. Invoking the 
radial velocity mapping in the  Galaxy~\citep{Vallee}, it can be seen that the
system velocity for LS\,III+52\degr24 (galactic coordinates l$\approx$100\degr, b$\approx-2\degr$)
and its large distance are in good agreement with the fact that the star belongs 
to the space beyond the Scutum–Crux arm.

    There are other objects in the family of hot post-AGB stars with similarly 
high velocities. For example, a significant  Vr=$-124.2\pm0.4$\,km/s was  determined
by~\citet{Ikon2020} for the post-AGB star LS\,5112  (IRAS\,18379$-$1707). The combination
of fundamental parameters and spectral features of LS~5112 obtained by~\citet{Ikon2020}
allows us to consider this star as the closest analog of LS\,III+52\degr24. We regard
the revealed excess of helium and CNO elements as an important result of these authors,
which directly  indicates the post-AGB evolution stage and the effectiveness of the
third dredge-up that took place. Unfortunately, it is not possible to compare the behavior
of the  spectral features of LS~5112 and LS\,III+52\degr24 over time, since the LS~5112
spectrum was studied by~\citet{Ikon2020}  based on a single
 observation. A related object is the hot post-AGB star  V886\,Her
(=IRAS\,18062+2410), for  which~\citep{IRAS18062} found a
photometric variability of the same amplitude, identified a set of forbidden 
emissions, as well as HeI wind components.  The closest relative of this star is 
the hot B-supergiant  LS\,II+34\degr26 (=\,V1853\,Cyg), in the spectrum of
 which~\citet{V1853Cyg} identified many envelope emissions
and recorded the systemic velocity of approximately $-49\pm5$\,km/s.

As noted in the Introduction,  the high-resolution optical spectrum of
LS\,III+52\degr24 was studied   earlier  in detail  by~\citet{Sarkar2012}.
These authors were the first to  determine its fundamental parameters, the chemical
composition of its atmosphere; they also found a high radial velocity from absorption 
lines, Vr=$-148.31\pm 0.60$\,km/s, and documented the status of the star as an
O-rich post-AGB star. But that study was also based on a single observation. 
Apparently, the time behavior of the optical spectrum in the family of hot 
post-AGB stars has been studied on the basis of high spectral resolution 
observations only for IRAS\,01005+7910~\citep{IRAS01005a, IRAS01005b} so far,
which located slightly closer in the Galaxy according to its reliable parallax 
$\pi=0.2414\pm0.0176$\,mas from Gaia DR3.

    We should note that the abovementioned features of the optical spectrum of 
LS\,III+52\degr24 (powerful emissions of HI and HeI lines with variability in
profiles, the presence of forbidden emissions of light metal ions), as well as its 
position near the plane of the Galaxy (galactic latitude b=$-1.96\degr$) allow us 
to suspect that this star belongs to the family of supergiants with the B[e] phenomenon, 
the principal features of the spectra of which were provided by~\citet{Lamers}.
A good example of a supergiant with the B[e] phenomenon is MWC\,17, a hot star several 
kiloparsecs away in the system of the source IRAS\,01441+6026 near the Galactic plane. 
As shown by~\citet{mimicry}, the optical spectrum of MWC\,17
contains powerful HI emissions and is rich with intense forbidden and permitted metal
emissions, while stellar absorptions are absent completely, with the exception of 
interstellar absorptions of the NaI D-lines and DIBs. However, the total set of 
available data for  LS\,III+52\degr24 (low absolute luminosity, chemical
composition features according to the data by~\citet{Sarkar2012}, and
high radial velocity) corresponds to the status of a hot post-AGB star. 
Thus, the spectrum of the supergiant LS\,III+52\degr24 is an example of spectral
mimicry of supergiants. This phenomenon  was previously considered in more detail 
by~\citet{mimicry}.

    The powerful H$\alpha$ emission in the spectrum of LS\,III+52\degr24
is 65--77 times higher than the level of the local continuum. Such a powerful 
emission in H$\alpha$ is an unique phenomenon for low-mass supergiants. As
follows from the papers by~\citet{V1853Cyg},  \citet{IRAS01005a, IRAS01005b} and
\citet{Ikon2020}, the H$\alpha$ emission intensity in the spectra of the closest
analogs, hot central post-AGB stars in the IR-source systems
IRAS\,01005+7910, IRAS\,18062+241, and IRAS\,18379$-$1707 is an order of magnitude lower.
Even in the spectra of  supergiants with the B[e] phenomenon, the H$\alpha$ emission is also significantly
lower (see the examples of profiles in the spectra of supergiants with the B[e] 
phenomenon in papers by~\citet{MWC17, IRAS00470, VES}). Such a strong H$\alpha$ emission can rather
be seen in the spectra of stars of extremely high luminosity, for example, in
the LBV spectra. However, even the spectrum of star No12 in the Cyg\,OB2 association, 
which is a well-known candidate for LBV,  the H$\alpha$ intensity
many times lower than that observed in the spectrum of LS\,III+52\degr24,
the luminosity of which is much lower. Apparently, this phenomenon is due to 
the significant contribution of the circumstellar gaseous medium and is related 
to the problem of spectral mimicry of supergiants~\citep{mimicry}. The absence
of an excess flux in the near-IR range and the high systemic velocity,
Vsys$\approx -150$\,km/s determined by us confirm the conclusion by~\citet{Sarkar2012}
that  LS\,III+52\degr24  belongs to the type
of low-mass supergiants at the post-AGB stage approaching the planetary nebulae.

\vspace{0.5cm}
    {\bf The main new results} obtained from multiple observations of the B-supergiant
LS\,III+52\degr24  in a wide wavelength range in 2010--2021 are as follows:
\begin{itemize}
\item reliable record of the LS\,III+52\degr24 systemic velocity from stationary
emissions in its spectrum: Vsys=$-149.6\pm 0.7$\,km/s;

\item conclusion about the significant distance to the star, d$\approx$5.3\,kpc;

\item detection of the velocity variability and stratification in the extended
atmosphere and the unhomogeneous envelope.  The position of wind absorption changes in the
interval from $-270$ to $-290$\,km/s. The wind speed reaches 150\,km/s;

\item detection of the time variability of the radial velocity based on 
the positions of photospheric absorptions of NII and OII ions in the range from
 $-127.2$ to $-178.3$\,km/s, which indicates the presence of a component or 
pulsations in the atmosphere;

\item discovery of the variability of the profile of the oxygen IR triplet OI\,7775\,\AA{} 
due to the appearance of unstable emission.
\end{itemize}

\vspace{0.5cm}
     It is obvious that finding the cause and determining the parameters of the 
detected variability in the radial velocity and line profiles will require further 
spectral  monitoring of  LS\,III+52\degr24 with a high spectral  resolution.

\section*{Acknowledges}

     V.G.K., who performed the analysis of the spectra and
 kinematic data for the  LS\,III+52\degr24 system, thanks the
 Russian Science Foundation for support (grant no.\,22-22-00043).
 The observations with telescopes of the Special Astrophysical Observatory, 
Russian Academy of Sciences, were supported by the Ministry of Science and Higher
 Education of the Russian Federation (agreement no.\,05.619.21.0016, project 
id.\,RFMEFI61919X0016). The study used the SIMBAD, VALD, SAO/NASA
 ADS, and Gaia DR3 astronomical databases.


\newpage

\begin{thebibliography}{99}

\bibitem[{Arkhipova} et al (2001a)]{IRAS18062} V.P.~Arkhipova,  V.G.~Klochkova, G.V.~Sokol, \alet\  \textbf{27}, 99  (2001a).

\bibitem[{Arkhipova}  et al (2001b)]{V1853Cyg} V.P.~Arkhipova,  N.~P.~Ikonnikova, R.~I,~Noskova, G.~V.~Komissarova, V.~G.~Klochkova, V.~F.~Esipov, \alet\ \textbf{27},  841 (2001b).

\bibitem[{Arkhipova} et al (2013)]{Arkh2013}  V.P.~Arkhipova,  M.A.~Burlak, V.A.~Esipov, N.P.~Ikonnikova, G.V.~Komissarova, \alet\  \textbf{39}, 619 (2013).

\bibitem[{Bailer-Jones} et al (2021)]{Bailer} C.A.L.~Bailer-Jones, J.~Rybizki, M.~Fouesneau, M.~Demleitner, R.~Andrae, \aj\
         \textbf{161} (3) 147 (2021).

\bibitem[{Bl\"ocker} (1995)]{Block1995} T.~Bl\"ocker, \aap\ \textbf{297}, 727 (1995).

\bibitem[{Di~Criscienzo} et al (2016)]{Criscienzo}  M.~Di~Criscienzo, P.~Ventura, D.A~Garc\`ia-Hern\`andez,
          F.~ Dell\'{}Agli,  M.~Castellani, P.M.~Marrese, S.~Marinoni, G.~Giuffrida, O.~Zamora, \mnras\ \textbf{462},  395 (2016).

\bibitem[{Garc\`ia-Lario}  et al (1997)]{Garcia1997}  P.~Garc\`ia-Lario,  M.~Parthasarathy, D.~de~Martino, L.~Sanz~Fernandez~de~Cordoba, R.~Monier, A.~Manchado, S.R.~Pottasch,    \aap\ \textbf{326}, 11037 (1997).


\bibitem[{Gauba} et al (2003)]{Gauba} G.~Gauba, M.~Parthasarathy,  B.~Kumar,  R.K.S.~Yadav, R.~Sagar, \aap\ \textbf{404}, 305  (2003).

\bibitem[{Gledhill} and {Forde}(2015)]{Gledhill}  T.M.~Gledhill, K.P.~Forde, \mnras\ \textbf{447}, 1080 (2015).


\bibitem[{Green} et al (2019)]{Green}  G.M.~Green, E.~Schlafly,  C.~Zucker, J.S.~Speagle, D.~Finkbeiner, \apj\ \textbf{887},
         (1) id.93 (2019).

\bibitem[{Hardorp} et al (1964)]{Hardorp} J.~Hardorp, I.~Theile,  H.H.~Vogt,  Hamburger Sternw., Warner \& Swasey Obs.,
   \textbf{3}, 0, (1964).

\bibitem[{Herwig} (2005)]{Herwig2005} F.~Herwig,  \araa\  \textbf{43}, 435 (2005).


\bibitem[{Hrivnak} et al (2009)]{Hrivnak} B.J.~Hrivnak, K.~Volk,  S.~Kwok, \apj\  \textbf{694}, 1147 (2009).

\bibitem[{Oudmaijer} et al (1992)]{Oudm1992} R.D.~Oudmaijer, W.E.C.J.~van~der~Veen, L.~B.~F.~M.~Waters, et al
                 \aas\ \textbf{96}, 625 (1992).

\bibitem[{Oudmaijer}(1996)]{Oudm1996} R.D.~Oudmaijer,  \aap\ \textbf{306}, 823 (1996).


\bibitem[{Ikonnikova} et al (2020)]{Ikon2020}  N.P.~Ikonnikova,  M.~Parthasarathy, A.V.~Dodin, S.~Hubrig,  G.~Sarkar,   \mnras\ \textbf{491}, 4828 (2020)

\bibitem[{Klochkova} (1999)]{rev1}  V.G.~Klochkova,  \bsao\ \textbf{44}, 5 (1999).

\bibitem[{Klochkova} (2014)]{Envelop}  V.G.~Klochkova,  \ab\  \textbf{69}, 279 (2014).

\bibitem[{Klochkova} (2019)]{rev2}  V.G.~Klochkova,  \ab\  \textbf{74}, 475 (2019).


\bibitem[{Klochkova} and {Chentsov}(2016)]{MWC17}  V.G.~Klochkova \& E.L.~Chentsov,  \ab\  \textbf{71}, 33 (2016).

\bibitem[{Klochkova} and {Chentsov}(2018)]{mimicry}  V.G.~Klochkova, E.L.~Chentsov,  \arep\ \textbf{62}, (1) 19 (2018).

\bibitem[{Klochkova} et al (2002)]{IRAS01005a}  V.G.~Klochkova,  M.V.~Yushkin,  A.S.~Miroshnichenko, V.E.~Panchuk, K.S.~Bjorkman, \aap\ \textbf{392}, 143 (2002).

\bibitem[{Klochkova} et al (2010)]{IRAS22223} V.G.~Klochkova,  V.~E.~Panchuk, N.~S.~Tavolzhanskaya, \arep\  \textbf{54}, 234 (2010).

\bibitem[{Klochkova} et al (2014)]{IRAS01005b}  V.G.~Klochkova,  E.L.~Chentsov,  V.E.~Panchuk, E.G.~Sendzikas, M.V.~Yushkin,  \ab\ \textbf{69}, 439 (2014).

\bibitem[{Kos} and {Zwitter}(2013)]{Kos}  J.~Kos \& T.~Zwitter, \apj\  \textbf{774}, 72 (2013).

\bibitem[{Kwok} (1993)]{Kwok}  S.~Kwok,  \araa\ \textbf{31}, 63   (1993).


\bibitem[{Lamers} et al (1998)]{Lamers}  H.J.G.L.M.~Lamers, F.J.~Zickgraf, D.~de~Winter, L.~Houziaux, J.~Zorec,   \aap\
    \textbf{340},  117  (1998).

\bibitem[{Liu} et al (2018)]{Liu}  N.~Liu, R.~Gallino, S.~Bisterzo,  A.M.~Davis, R.~Trappitsch, L.R.~Nittler,  \apj\ \textbf{865}, 112 (2018).

\bibitem[{Luna} et al (2008)]{Luna}  R.~Luna, R.N.L.J.~Cox, M.A.~Satorre, D.A.~Garc\'ia Hern\'andez,  O.~Su\'arez,
            P.~Garc\`a Lario, \aap\  \textbf{480}, (1)  133  (2008).

\bibitem[{Mello} et al (2012)]{Mello2012}  M.~Mello, S.~Dafton,  C.B.~Pereira, I.~Hubeny, \aap\  \textbf{543}  A11 (2012).

\bibitem[{Miroshnichenko} (2007)]{Mir2007} A.S.~Miroshnichenko, \apj\  \textbf{667}, 497 (2007).

\bibitem[{Miroshnichenko} et al (2009)]{IRAS00470} A.S.~Miroshnichenko, E.L.~Chentsov, V.G.~Klochkova, S.V.~Zharikov,
  K.N.~Grankin, A.V.~Kusakin, T.L.~Gandet, G.~Klingenberg, S.~Kildahl, R.J.~Rudy, D.K.~Lynch,
  C.C.~Venturini, S.~Mazuk,  R.C.~Puetter, R.B.~Perry, A.C.~Carciofi,  K.S.~Bjorkman,
  R.O.~Gray,  S.~Bernabei, V.F.~Polcaro,   R.F.~Viotti, L.~Norci, \apj\ \textbf{700}, 209 (2009).

\bibitem[{Miroshnichenko} et al (2021)]{VES} A.S.~Miroshnichenko,  V.G.~Klochkova, E.L.~Chentsov, V.E.~Panchuk,
    M.V.~Yushkin  and N.~Manset, \mnras\ \textbf{507}, 879 (2021).

\bibitem[{Pakhomov} et al (2019)]{VALD2}  Yu.V.~Pakhomov, T.A.~Ryabchikova, N.E.~Piskunov,  \arep\ \textbf{63},  1010 (2019).

\bibitem[{Panchuk} et al (2017)]{NES}   V.E.~Panchuk, V.G.~Klochkova,  M.V.~Yushkin, \arep\  \textbf{61}, 820  (2017).

\bibitem[{Parthasarathy} et al (1993)]{Partha1993}  M.~Parthasarathy,  P.~Garcia-Lario,  S.R.~Pottasch, A.~Manchado, J.~Clavel, D.~de~Martino,  G.C.M.~van~de~Steene, K.C.~Sahu,  \aap\, \textbf{267},  L19  (1993).

\bibitem[{Parthasarathy} et al (2020)]{Partha2020}   M.~Parthasarathy, T.~Matsuno, W.~Aoki, \pasj\, \textbf{72} (6) 99  (2020).

\bibitem[{Pottash} \& {Parthasarathy}(1998)]{Pottash1998}  S.R.~Pottash, M.~Parthasarathy, \aap\ \textbf{192}, 182   (1998).


\bibitem[{Ryabchikova}  et al (2015)]{VALD1} T.~Ryabchikova, N.~Piskunov, R.L.~Kurucz, H.C.~Stempels,  U.~Heiter, Yu.~Pakhomov, P.S.~Barklem, Physica Scripta, \textbf{90}, (5) article id.\,054005 (2015).


\bibitem[{Sarkar} et al (2005)]{Sarkar2005}  G.~Sarkar, M.~Parthasarathy, B.E.~Reddy, \aap\ \textbf{431}, 1007 (2005).

\bibitem[{Sarkar} et al (2012)]{Sarkar2012} G.~Sarkar, D.~A.~Garc\`ia-Hern\`andez, M.~Parthasarathy, A.~Manchado,
    P.~Garcia-Lario, Y.~Takeda,  \mnras\ \textbf{421}, 679 (2012).

\bibitem[{Su\`arez} et al (2006)]{Suarez} O.~Su\`arez, P.~Garc\`ia-Lario, A.~Manchado,  M.~Manteiga, A.~Ulla,  S.R.~Pottasch,
         \aap\ \textbf{458}, 173 (2006).

\bibitem[{Vallee} (2008)]{Vallee}  J.P.~Vallee, \aj\  \textbf{135}, 1310   (2008).


\bibitem[{Yushkin} and {Klochkova}(2005)]{MIDAS}  M.V.~Yushkin and V.G.~Klochkova, Special Astrophysical Observatory,
           Preprint No.\,206,  (2005).


\end{thebibliography}
\end{document}